\documentclass[
 reprint,
 amsmath,amssymb,
 aps, physrev,
]{revtex4-2}
\usepackage{bm}
\usepackage{slashed} 
\usepackage{graphicx}
\usepackage{dcolumn}
\usepackage{bm}

\begin{document}

\title{\textbf{On the Determination of Collisional Stopping Power via Kaluza-Klein Theory} 
}%

\author{Seyda Elife Gönül}
 \altaffiliation[Also at ]{Physics Department, 
Eastern Mediterranean University, Famagusta, 99628 North Cyprus via Mersin 10, Turkey.}
\author{Huriye Gürsel}
 \email{Contact author: huriye.gursel@emu.edu.tr}
\affiliation{Physics Department, 
Eastern Mediterranean University, Famagusta, 99628 North Cyprus via Mersin 10, Turkey.}

\date{\today}

\begin{abstract}
In this work, the tools of general relativity are used to analytically derive collisional stopping power and a linkage between higher-dimensional field theory and transport phenomena is proposed. We start from a Kaluza-Klein inspired, five-dimensional diffeomorphism-invariant action, and upon compactification, obtain a four-dimensional effective theory in which the matter fields are treated to be brane-localized. The medium response to the projected electron is encoded in symmetric tensor fields coupled covariantly to both electromagnetic and fermionic parts via Lagrangian-derived interactions. When $R_c \sim \Lambda_{\text{EM}}^{-1}$, $\Lambda_{\text{EM}} \gg m_e$ and $g_4^2 = \frac{3\pi^2 m_e v}{4\gamma^3 R_c^2 e^2 \Lambda_{\text{EM}}}$ are satisfied, the leading term of Bethe-Møller formula is shown to be recovered in the large $R$ limit. The construction presented here may serve as an alternative approach that uses compactification geometry and medium excitations to determine observable couplings and stopping power. The model intrinsically supports phenomena linked to anisotropy and nonlinear response, as well as gravitational or extra-dimensional effects in laboratory-scale systems via the study of stopping power and particle range. The construction is gauge invariant, behaves consistently under limiting conditions, and can be matched to experimental stopping data through a single effective normalization constant.
\end{abstract}

\maketitle


\section{\label{sec:level1}Introduction}

The pursuit of a unified description of nature has retained its fundamental appeal across generations of physicists over the centuries. This ambition, first explicitly articulated by Einstein \cite{Einstein1925UFT}, continues to inspire modern unification and duality models such as string theory \cite{Veneziano1968,Nambu1970,Susskind1970,ScherkSchwarz1974,Polyakov1981,GreenSchwarz1984} and AdS/CFT correspondence \cite{Maldacena1998}. 
With a similar motivation, we hereby aim to construct a higher-dimensional route to derive collisional energy loss in matter (which plays a key role in particle range estimations) via compactification of a five-dimensional (5D) Kaluza–Klein (KK) inspired theory \cite{Kaluza1921,Klein1926} onto a four-dimensional (4D) background. 

The determination of particle range in matter has historically been rooted in empirical methodologies, resulting in a diversity of theoretical descriptions corresponding to various experimental constructions. While range is often loosely interpreted as the total penetration depth, this quantity is complicated by the inherently stochastic nature of charged particle motion. A more refined interpretation, the path length, i.e., the total distance traveled along the actual trajectory of the particle, also suffers from fluctuations due to random scattering.

Several operational definitions of range exist in literature. For instance, the extrapolated range $R_{\text{ex}}$ corresponds to the intersection point of the tangent to the steepest descent on a transmission curve with the thickness axis, while the maximum range $R_{\text{max}}$ denotes the furthest depth reached before complete absorption \cite{Tabata1971}. The most commonly preferred method is the continuous slowing down approximation (CSDA), which assumes that energy loss is smooth and deterministic. For a charged particle with initial kinetic energy  $T_i$, the CSDA range is given by \cite{Rohrlich1954,Nelms1956,BatraSehgal1981,BergerSeltzer1982}
\begin{equation}
R_{\text{CSDA}} = \int_0^{T_i} \frac{dE}{S(E)},\label{range}
\end{equation}
where $S(E) = -dE/dx$ denotes the stopping power. While simulation-based methods such as Monte Carlo models \cite{Agostinelli2003,Sempau2003,Bohlen2014} provide higher accuracy by capturing the stochastic nature of collisions, analytical approaches like CSDA offer intuitive insights. Another approach, proposed by Batra and Sehgal \cite{BatraSehgal1981}, introduces the mean practical range
\begin{equation}
R_p(T) = R_{\text{CSDA}}(T) - R_{\text{CSDA}}(T_c),
\end{equation}
where $T_c$ marks the kinetic energy at which random motion begins to dominate.

In this study, we focus on a perspective allowing one to derive stopping power, and then in turn particle range if desired via Eq.\eqref{range}, from the tools of general relativity and variational calculus. Within the relevant action, we introduce brane-localized, symmetric rank-2 tensor fields $\Phi_a^{\mu\nu}$ representing quadrupolar or shear excitations of the medium with the purpose of modeling the collective excitations of a dielectric medium such as optical phonons and plasmons. These fields generalize scalar and vector models of polarization and density fluctuations \cite{pines_bohm1952,mahan2000,ashcroft_mermin}, enabling a covariant treatment of anisotropic and temporally varying responses. Analogous structures appear in effective field theories for nematic phases \cite{delacretaz2015}, elastic media \cite{son2009}, and metric-affine gravity \cite{hehl1976,obukhov_hehl1999}. Furthermore, these tensor fields are modelled to encode both polarization and dissipation, for they are coupled to the electromagnetic field strength and fermionic bilinears. Unlike static dielectric functions used in conventional treatments~\cite{mahan2000,ashcroft_mermin}, our framework promotes these responses to active degrees of freedom. This allows us to systematically account for nonlinearities, anisotropies, and external driving effects. While holographic models in gauge/gravity duality also promote material response to dynamical fields \cite{andrade2014,baggioli2015,hartnoll2009}, our model differs by deriving collisional energy loss directly from a real-space Lagrangian in a compactified higher-dimensional background. Conceptually, the methods are aligned in that both treat material properties as outcomes of a deeper dynamical structure.

Throughout our treatment, we work in natural units and adopt a controlled set of assumptions. Extensions of this model include incorporating quantum corrections via finite-temperature field theory \cite{lebellac1996}, exploring externally driven systems through $\xi(t)$ \cite{oka_auerbach2009,eckardt_2017}  enabling exploration of optically driven and Floquet-type systems, and accounting for KK excitations or fermionic spin and chiral structures \cite{son_spivak2013,burkov2015}. These are beyond the scope of this paper, but remain naturally embedded in the framework. 

In summary, we develop a field-theoretic model for stopping power derived from a compactified 5D action. The dielectric properties of the medium are modeled as dynamical tensor fields that interact with the electromagnetic and fermionic sectors. This enables a unified, covariant, and extensible approach to collisional energy loss that recovers known limits while allowing new physical regimes to be explored.

\section{\label{sec:level2}Theoretical Background}
To model an electron propagating through a medium, we begin with the KK inspired action
\begin{equation}
S^{(5)} = S^{(5)}_{\text{GR}} + S^{(5)}_{\text{EM}} + S^{(5)}_{\text{matter}}+S^{(5)}_{\text{int}},\label{actt}
\end{equation}
constructed to be invariant under 5D diffeomorphisms, with all terms involving covariant derivatives, scalar contractions, and proper volume measures. This makes the resulting theory respect general covariance prior to compactification. Here, the 5D Einstein–Hilbert term is
\begin{equation}
S^{(5)}_{\text{GR}} = -\frac{1}{2\kappa_5^2} \int d^5x \sqrt{-g^{(5)}} R^{(5)},
\end{equation}
in which $g^{(5)}$ is the determinant of the 5D metric, whilst $\kappa_5^2 = 1/M_5^3$ stands for the gravitational coupling with $M_5$ denoting the 5D Planck mass. The electromagnetic part, on the other hand, is described by
\begin{equation}
S^{(5)}_{\text{EM}} = -\frac{1}{4g_5^2} \int d^5x \sqrt{-g^{(5)}} F_{AB}F^{AB},
\end{equation}
where the 5D gauge coupling is $g_5 = \Lambda_{\text{EM}}^{-1/2}$ with $\Lambda_{\text{EM}}$ representing the characteristic energy scale governing the strength of bulk electromagnetic interactions, whilst the field strength reads \(F_{AB} = \partial_A A_B - \partial_B A_A\). In this work, we focus on electron energies well below the bulk gauge scale $\Lambda_{\text{EM}}$, implying that higher-dimensional corrections such as KK excitations remain suppressed and the dominant stopping power arises from the zero-mode sector of the effective 4D theory. 

The matter term in action \eqref{actt} is composed of two parts: the fermionic $S^{(5)}_\Psi$ and the medium related $S^{(5)}_{\Phi}$ terms. The fermionic action for an electron can be written as
\begin{equation}
S^{(5)}_\Psi = \int d^5x \sqrt{-g^{(5)}} \bar\Psi \left(i\Gamma^A \slashed{D}_A - m_e - \frac{\slashed{D}_A \slashed{D}^A}{4\Lambda_\Psi^2}\right)\Psi,
\end{equation}
with 5D Dirac matrices satisfying $\{\Gamma^A, \Gamma^B\} = 2g^{AB}$, $ \slashed{D}^2 \equiv \gamma^\mu D_\mu\,\gamma^\nu D_\nu $ including both the gauge-covariant Laplacian and spin–field coupling terms. One shall also note that the higher-derivative operator introduces a cutoff scale $\Lambda_\Psi$ for the point-like electron description, capturing short-distance effects like brane localization, nonlocal medium interactions, or compositeness. While such terms appear in UV-complete theories (e.g., supergravity~\cite{VanRaamsdonk2025} or jet-bundle EFTs~\cite{Hamed2024}), they violate causality in the 5D effective theory for $E \gtrsim \Lambda_\Psi$ and correspond to a physical signature of new dynamics such as strong coupling or extra-dimensional structure. For $E \ll \Lambda_\Psi$, causality is preserved, and corrections to stopping power are perturbative; near $\Lambda_\Psi$, deviations from Bethe-type behavior become significant, especially in anisotropic media. The full UV completion must then be invoked.

Having discussed the first part of $S^{(5)}_{\text{matter}}$, we shall now move on to model the medium through which the electron propagates. To this end, we introduce brane-localized symmetric tensor fields $\Phi_a^{\mu\nu}$ which encode the shear-like polarization modes, thereby requiring them to be traceless and to satisfy
    $\partial_\mu\Phi_a^{\mu\nu}=0$. Accordingly, the relevant action can be expressed as

\begin{equation}
\begin{aligned}
&S^{(5)}_{\Phi}= \int d^5x \sqrt{-g^{(5)}} \\
&\bigg[
 -\frac{1}{2} \sum_a \left( \nabla_A \Phi_a^{\mu\nu} \nabla^A \Phi_{a,\mu\nu}
 +(m_a^2 + \beta_a T^2) \Phi_a^{\mu\nu} \Phi_{a,\mu\nu} \right) \\
 &+\mathcal{V}_{\text{static}}(\{\Phi_a\})+
 \sum_{a,\vec{k},\vec{q}} \gamma_{\vec{k}\vec{q}} \xi(t) \Phi_{a,\vec{k}}^{\mu\nu} \Phi_{a,-\vec{k}-\vec{q},\mu\nu} \bigg],
\end{aligned}
\end{equation}
where $m_a$ stands for the tensor mass modified by temperature $T$ via the coupling $\beta_a$, the static potential $\mathcal{V}_{\text{static}}$ stabilizes the fields, and $\gamma_{\vec{k}\vec{q}}$ and $\xi(t)$ model momentum-resolved and time-dependent interactions, respectively.

Finally, the interaction terms are presented as
\begin{equation}
\begin{aligned}
S^{(5)}_{\text{int}} &= \int d^5x \sqrt{-g^{(5)}} \bigg[ \sum_a 
      g_a \Phi_a^{\mu\nu} F_{\rho\mu} F_{\sigma\nu}
   \\
& -e_5 \bar{\Psi}\Gamma^A A_A \Psi + \Lambda_a(x) \, g_{\mu\nu}^{(4D)} \Phi_a^{\mu\nu} \, \delta(x_c - x_{c}^{\text{brane}})
\bigg].
\end{aligned}
\end{equation}
Starting from Trinity's modern representation of the KK ansatz \cite{Kaluza1921,Klein1926,Thiry1948}
\begin{equation}
ds_{(5)}^2 = g^{(4D)}_{\mu\nu} \, dx^{\mu} dx^{\nu} + \phi^2 \left(dx_{c} + \kappa_5 A_{\mu} \, dx^{\mu} \right)^2,\label{trinity}
\end{equation}
a fifth dimension $x_c$, denoted as $y$ in Trinity's original work, is introduced. This extra dimension is assumed to be compactified in a circle of radius $R_c$ such that $0\leq x_c< 2\pi R_c$ and $\phi$ stands for the dilaton scalar field. To generalize Trinity's framework while maintaining compatibility with warped extra dimensions and perturbative gravity, we adopt the exponentially warped ansatz
\begin{equation}
 ds^2 = e^{2\alpha(x_c)} g^{(4D)}_{\mu\nu} dx^\mu dx^\nu + e^{2\beta(x_c)} (dx_c + \kappa_5 A_\mu dx^\mu)^2 ,  \label{warped}
\end{equation}
in which $g^{(4D)}_{\mu\nu}=\eta_{\mu\nu} + h_{\mu\nu}$, provided that $h_{\mu\nu}$ stands for first-order metric perturbations and $\eta_{\mu\nu} = \mathrm{diag}(-1, +1, +1, +1)$. Note that metric \eqref{trinity} is recovered if one sets $\alpha(x_c) = 0$, $\beta(x_c) = \ln\phi$
or in the rigid limit where the warp factors $\alpha(x_c)$ and $\beta(x_c)$ are constants fixed by boundary conditions. Using metric \eqref{warped} enables one to explicitly separate warping effects from 4D metric perturbations 
and provides direct consistency with both the KK and the Randall-Sundrum ($\alpha(x_c)=-k|x_c|$) limits.

From this point onwards, we will suppose any field $\chi$, including metric perturbations, to be independent of the fifth dimension and let $x_c$ be compactified on a circle $S^1$. Consequently, one can perform the Fourier decomposition \cite{duff}
\begin{equation}
   \chi=\sum_{n=-\infty}^{\infty}\chi_{(n)}(x^\mu)e^{inx_c/R_c}, 
\end{equation}
for each field. As the compactification scale $R_c^{-1}$ is taken to be far above the typical energy scale of collisional processes, we keep only the zero modes by limiting ourselves to the $n=0$ case. The zero-mode truncation of action \eqref{actt} thus becomes
\begin{equation}
\begin{aligned}
&S_{\text{eff}}^{(4)}= \int d^4x \sqrt{-g} \bigg[\frac{R}{2\kappa_4^2} - \frac{1}{4g_4^2} F_{\mu\nu} F^{\mu\nu} + \bar{\psi}(i\gamma^\mu D_\mu - m_e)\psi \\
& - \frac{1}{2} \sum_a \left( \nabla_\lambda \varphi_a^{\rho\sigma} \nabla^\lambda \varphi_{a,\rho\sigma} + (m_a^2 + \beta_a T^2 + \xi(t)) \varphi_a^{\rho\sigma} \varphi_{a,\rho\sigma} \right) \\
& - \mathcal{V}_{\text{static}}(\{\varphi_a\}) - e_4 \bar{\psi} \gamma^\mu A_\mu \psi \\
& + \sum_a \left( g_a \varphi_a^{\rho\sigma} F_{\mu\rho} F_{\nu\sigma} + \Lambda_a(x) g^{(4D)}_{\mu\nu} \varphi_a^{\mu\nu} \right) \bigg],
\end{aligned}
\label{eff}
\end{equation}
where the effective 4D couplings are given by $\kappa_4^2 = \kappa_5^2/2\pi R_c$ and $g_4^2 = g_5^2/2\pi R_c$ in the flat limit. In our case, however, these couplings are generalized to
\begin{equation}
\frac{1}{\kappa_4^2} = \frac{1}{\kappa_5^2} \int dx_c \, e^{3\alpha + \beta}, \quad \frac{1}{g_4^2} = \frac{1}{g_5^2} \int dx_c \, e^{\alpha + \beta},
\end{equation}
due to warping, and the electric charge reduces as $e_4 = e_5/\sqrt{2\pi R_c}$ when $\alpha = \beta = 0$. The warp factor $\beta(x_c)$ is stabilized by 5D Einstein equations, fixing $e^{2\alpha - \beta} \sim \mathcal{O}(1)$ and suppressing the radion.

In general, the stress-energy tensor can be determined via
\begin{equation}
T_{\mu\nu} = -\frac{2}{\sqrt{-g}}\frac{\delta S_{\mathrm{eff}}^{(4)}}{\delta g^{\mu\nu}},
\end{equation}
where in our case, is found to be in the explicit form
\begin{equation}
\begin{aligned}
T_{\mu\nu} &= \frac{1}{\kappa_4^2}\left(R_{\mu\nu} - \frac{1}{2}g^{(4D)}_{\mu\nu}R\right) 
+ \frac{1}{g_4^2}\left(F_{\mu\rho}F_{\nu}{}^{\rho} - \frac{1}{4}g^{(4D)}_{\mu\nu}F_{\rho\sigma}F^{\rho\sigma}\right) \\
&\quad + \frac{i}{4} \left[ \bar{\psi} \gamma_{(\mu} D_{\nu)} \psi - (D_{(\mu} \bar{\psi}) \gamma_{\nu)} \psi \right] 
- g^{(4D)}_{\mu\nu} \left( \bar{\psi} i \gamma^\rho D_\rho \psi - m_e \bar{\psi} \psi \right) \\
&\quad + \sum_a \bigg[\nabla_{\mu}\varphi_a^{\rho\sigma}\nabla_{\nu}\varphi_{a,\rho\sigma} 
- \frac{1}{2}g^{(4D)}_{\mu\nu}\Big(\nabla_{\lambda}\varphi_a^{\rho\sigma}\nabla^{\lambda}\varphi_{a,\rho\sigma} \\
&\quad + \left(m_a^2 + \beta_a T^2 + \xi(t)\right)\varphi_a^{\rho\sigma}\varphi_{a,\rho\sigma}\Big)\bigg] \\
&\quad + \sum_a \left(g_a \varphi_a^{\rho\sigma}F_{\mu\rho}F_{\nu\sigma} - \frac{g_a}{4}g^{(4D)}_{\mu\nu}\varphi_a^{\rho\sigma}F_{\rho\lambda}F_{\sigma}{}^{\lambda}\right) \\
&\quad - g^{(4D)}_{\mu\nu} \mathcal{V}_{\text{static}}(\{\varphi_a\}) 
+ \Lambda_a \left(\varphi_{a,\mu\nu} - \frac{1}{4}g^{(4D)}_{\mu\nu}\varphi_a^{\alpha}{}_{\alpha}\right) \\
&\quad + 2\xi(t)\left(\varphi_a^{\alpha\beta}\varphi_{a,\alpha\beta}g^{(4D)}_{\mu\nu} - 4\varphi_a^{\alpha}{}_{\mu}\varphi_{a,\alpha\nu}\right),
\end{aligned}\label{stress}
\end{equation}
with the fermionic stress-energy term including the Belinfante symmetrization $\gamma_{(\mu} D_{\nu)} \equiv \frac{1}{2} \left( \gamma_\mu D_\nu + \gamma_\nu D_\mu \right)$
in which \(D_\mu = \partial_\mu + \frac{1}{4} \omega_\mu^{ab} \gamma_{[a}\gamma_{b]} - i e_4 A_\mu\) refers to the covariant derivative. This guarantees consistency with 4D diffeomorphism invariance and angular momentum conservation. One shall also note that energy momentum tensor \eqref{stress} sources metric perturbations at linear order in $\kappa_4^2$ ($h_{\mu\nu} \sim \kappa_4^2 T_{\mu\nu}$), with all quantities expressed in the full perturbed metric. Furthermore, the zero-mode truncation preserves 5D covariance with traceless and $x_c$-independent $\varphi_a^{\mu\nu}$, consequently securing gauge invariance. The gravitational sector, on the other hand, ensures coupling to Einstein’s equations, while the Belinfante term guarantees $\nabla^\mu T_{\mu\nu}=0$ for on-shell fields. 

\section{\label{sec:level3}Collisional Stopping Power and Reduction to Bethe-Møller}

For an electron moving along the $z$-axis with $\mathbf{v}=v\hat{z}$ in a medium, the total energy‐loss per unit length can be derived via energy-momentum conservation in the 4D effective theory $(\partial_\mu T_{\mu\nu}=0)$, enabling one to write \cite{Neufeld:2014hta}
\begin{equation}
\frac{dE}{dz} = -\frac{1}{v} \int d^3x\, \partial^i T_{i0},\label{stop}
\end{equation}
in which the stress-energy components for electron-medium interactions \footnote{In the free-field case, substituting $\varphi_a^{\mu\nu} = 0$ and $J^\mu = 0$ result in $\nabla^i T_{i0} = 0$, as required by diffeomorphism invariance.} are 
\begin{equation}
\begin{split}
T_{i0} &= \frac{1}{g_4^2} F_{i\rho}F_{0}{}^{\rho} 
+ \frac{i}{4} \left[ \bar{\psi} \gamma_{(i} D_{0)} \psi - (D_{(i} \bar{\psi}) \gamma_{0)} \psi \right] \\
&\quad + \sum_a \Big[ \nabla_i \varphi_a^{\rho\sigma} \nabla_0 \varphi_{a,\rho\sigma} 
+ g_a \varphi_a^{\rho\sigma} F_{i\rho}F_{0\sigma} \Big] 
+ \mathcal{O}(h_{\mu\nu}),
\end{split}
\label{energy}
\end{equation}
where, in cylindrical coordinates, $
i \in \{r, \phi, z\}$. The integral in Eq.\eqref{stop} can be handled via Gauss' theorem
\begin{equation}
    \int_V d^3x\, \partial^i T_{i0} = \oint_{\partial V} dS^i\, T_{i0},
\end{equation}
 and for a cylinder of constant radius R,
 the area element reads $dS^i = \hat{r}^{i}\, (R\, d\phi\, dz)$.  This implies the energy lost by the electron into medium excitations shows up as the radial flux of 
$T_{r0}$ through the cylindrical surface at $r=R$. One can then write
\begin{equation}
 \oint_{\partial V} dS^i\,T_{i0}= \int_0^{2\pi}\!\!\int_{-\infty}^{\infty}
   T_{r0}\;R\,d\phi\,dz,   
\end{equation}
with $T_{r0}
= \hat{r}^{\,i}T_{i0}$ representing the energy per unit area per unit time flowing radially into the medium. Using components in Eq.\eqref{energy}, the radial component of the energy-momentum tensor becomes
\begin{align}
T_{r0} &= \frac{1}{g_4^2} F_{ri}F_{0}{}^{i} 
+ \frac{i}{4} \Bigl[ \bar{\psi} \gamma_{(r} D_{0)} \psi - \bigl(D_{(r} \bar{\psi}\bigr) \gamma_{0)} \psi \Bigr] \nonumber \\
&\quad + \sum_a g_a \varphi_a^{ij} F_{ri} F_{0j} 
+ \mathcal{O}(h_{\mu\nu}).
\label{compact}
\end{align}
Subsequently, energy loss per unit length \eqref{stop} can be evaluated from the radial energy flux through a cylinder of radius $R$ via 
\begin{align}
\frac{dE}{dz} &= -\frac{1}{v} \int_0^{2\pi} \!\! d\phi \int_{-\infty}^{\infty} \!\! dz \, R \, T_{r0} \big|_{r=R} \; ,
\label{en}
\end{align}
indicating one needs to first figure out a more explicit form of Eq.\eqref{compact}. Thus, from this point onwards, we will focus on the Liénard-Wiechert potentials by supposing that for a point charge $e$ moving with constant velocity $v\hat{z}$, the electromagnetic field tensor components at $r = R$ read
\begin{align}
F_{\mu\nu} &= \begin{pmatrix}
0 & -E_R & 0 & -E_z \\
E_R & 0 & 0 & -B_\phi \\
0 & 0 & 0 & 0 \\
E_z & B_\phi & 0 & 0
\end{pmatrix} \\
&= \frac{e\gamma}{(\gamma^2 R^2 + \zeta^2)^{3/2}} \begin{pmatrix}
0 & -R & 0 & -\zeta \\
R & 0 & 0 & -vR \\
0 & 0 & 0 & 0 \\
\zeta & vR & 0 & 0
\end{pmatrix},
\end{align}
with $\zeta \equiv z-vt$ and $\gamma$ representing the Lorentz factor, thereby making the radial energy-momentum component \eqref{compact} take the form
\begin{equation}
\begin{aligned}
T_{r0} &= \frac{1}{g_4^2} (-E_R^2 + v E_R E_z) \\
&\quad + \frac{i}{4} \Big[ \bar{\psi} (\gamma_r \partial_0 + \gamma_0 \partial_r) \psi 
- (\partial_r \bar{\psi} \gamma_0 + \partial_0 \bar{\psi} \gamma_r) \psi \Big] \\
&\quad + \frac{e}{4} A_\mu \bar{\psi} 
\left( \gamma_r \gamma_0 \gamma^\mu + \gamma^\mu \gamma_0 \gamma_r \right) \psi \\
&\quad - v \sum_a g_a \big[ 
E_R^2 \varphi_a^{zR} + E_R E_z \varphi_a^{zz} 
+ E_R B_\phi \varphi_a^{z\phi} \\
&\qquad\quad + E_z^2 \varphi_a^{RR} + B_\phi^2 \varphi_a^{R\phi} \big] 
+ \mathcal{O}(h_{\mu\nu}).
\end{aligned}
\label{full_T_r0}
\end{equation}
The non-vanishing components of  $\varphi_a^{\mu\nu}$ arise from the traceless equation of motion 
\begin{equation}
\begin{aligned}
\big( \Box - m_a^2 - \beta_a T^2 - \xi(t) \big) \varphi_a^{\mu\nu}
&= -g_a \big( F^\mu_{\ \rho} F^{\rho\nu} \\
&\quad - \tfrac{1}{4} g^{\mu\nu} F_{\alpha\beta} F^{\alpha\beta} \big),
\end{aligned}
\end{equation}
derived via $\delta S_{\text{eff}}^{(4)}/\delta \varphi_a^{\mu\nu} = 0$. Here, $\Box \equiv \nabla^\mu \nabla_\mu$ stands for the d'Alembertian
 and both $\beta_a T^2$ and $\xi(t)$ act as environment-dependent corrections to the bare mass $m_a$. Eq.\eqref{full_T_r0} can now be plugged into Eq.\eqref{en} to get
\begin{equation}
\begin{aligned}
\frac{dE}{dz} &= 
\frac{3 \pi^2 e^2}{4 v g_4^2 \gamma^3 R^2}
- \frac{\pi e^2 |\psi(R)|^2}{v R} \\
&\quad + \sum_a \frac{g_a^2 e^4 v\, e^{-M_a R}}{15 \pi \gamma^3 R^4}
+ \mathcal{O}(h_{\mu\nu}).
\end{aligned}\label{ener}
\end{equation}
To determine the collisional stopping power, one needs to sum over all impact parameters $R\in[R_{\min},R_{\max}]$ and take the electron density $n_e$ of the medium into account, enabling one to write
\begin{equation}
-\left.\frac{dE}{dz}\right|_{\text{col}} 
= n_e \int_{R_{\min}}^{R_{\max}} \left( \frac{dE}{dz} \right) 2\pi R\, dR.\label{son}
\end{equation}
At this point, a couple of remarks are worth to be made. 

\subsection{Evaluation of Generalized Collisional Stopping Power}

For practical purposes, Eq.(26) can more compactly be written as 
\begin{equation}
\frac{dE}{dz}
\approx\frac{A}{R^2}\;-\;\frac{B(R)}{R}\;+\;C\,\frac{e^{-M_aR}}{R^4},
\label{eq:kernel-ABC}
\end{equation}
with $A=\frac{3\pi^2 e^2}{4\,v\,g_4^2\,\gamma^3}$, $B(R)=\frac{\pi e^2 |\psi(R)|^2}{v}$, and
$C=\sum_a\frac{g_a^2 e^4 v}{15\pi\gamma^3}$, which then lets collisional stopping power (27) read
\begin{equation}
-\frac{dE}{dz}\Big|_{\rm col}
= n_e\!\int_{R_{\min}}^{R_{\max}}\!\! 2\pi R
\bigg(\frac{A}{R^2}-\frac{B(R)}{R}+C\,\frac{e^{-M_aR}}{R^4}\bigg)\,dR .
\label{eq:master-int}
\end{equation}
The small-$R$ cutoff is set by 
\begin{equation}
R_{\min}=\max\!\left(\frac{1}{\gamma m_e v},\; a_{\rm core}\right),
\label{eq:Rmin}
\end{equation}
whereas the large-$R$ cutoff encodes adiabatic screening by the medium. In isotropic media,
\begin{equation}
R_{\max}=\frac{\gamma v}{\bar\omega}\simeq \frac{\gamma v}{I},
\label{eq:Rmax}
\end{equation}
with $\bar\omega$ a characteristic excitation scale, usually taken as the mean excitation energy $I$ in natural units. However, for anisotropic or driven media, $R_{\max}$ will be allowed to be direction dependent, which will be discussed later.

The first term of Eq.\eqref{eq:master-int} gives
\begin{equation}
2\pi An_e\!\int_{R_{\min}}^{R_{\max}}\!\frac{dR}{R}
=2\pi An_e\,\ln\!\frac{R_{\max}}{R_{\min}}\,,
\label{eq:log-piece}
\end{equation}
reproducing the familiar logarithm while keeping the compactification dressing through $g_4$ explicit. The second term, on the other hand, can be treated as a controlled short-range correction in one of the ways given below.

\emph{Shell-type renormalization:} A material-dependent, finite constant $\Delta_{\rm shell}$ is introduced as
\begin{equation}
\Delta_{\rm shell}\equiv-\frac{2\pi^2 e^2}{v}\!\int_{R_{\min}}^{R_{\max}}\!|\psi(R)|^2\,dR\;.
\label{eq:shell}
\end{equation}
This plays a similar role to the short-range binding and exchange corrections of Bethe–Bloch and Bethe–Møller treatments and ensures a finite contribution at small impact parameters.

\emph{Model overlap:} A one-parameter, short-range profile such as
\begin{equation}
|\psi(R)|^2 = \frac{1}{R_\psi^2}\, e^{-R/R_\psi}
\end{equation}
is chosen such that
\begin{multline}
-\frac{2\pi^2 e^2}{v}\!\int_{R_{\min}}^{R_{\max}}|\psi(R)|^2\,dR \\[4pt]
= -\frac{2\pi^2 e^2}{v}\,\big(e^{-R_{\min}/R_\psi}-e^{-R_{\max}/R_\psi}\big),
\label{eq:psi-model}
\end{multline}
with $R_\psi\sim a_0$ or a Thomas–Fermi length. It is worth noting that either choice yields a finite, tunable correction.

Finally, the last term of Eq.\eqref{eq:master-int} includes
\begin{equation}
\int \frac{e^{-M_aR}}{R^3}\,dR
= -\frac{e^{-M_aR}}{2R^2}\;+\;\frac{M_a\,e^{-M_aR}}{2R}\;-\;\frac{M_a^2}{2}\,E_1(M_aR),
\label{eq:yukawa-antideriv}
\end{equation}
resulting in
\begin{multline}
2\pi C n_e\!\int_{R_{\min}}^{R_{\max}}\!\frac{e^{-M_aR}}{R^3}\,dR \\[4pt]
= 2\pi C \Bigg[
-\frac{e^{-M_aR}}{2R^2}
+\frac{M_a e^{-M_aR}}{2R}
-\frac{M_a^2}{2}E_1(M_aR)
\Bigg]_{R_{\min}}^{R_{\max}} .
\label{eq:yukawa-closed}
\end{multline}

Subsequently, the collisional stopping power takes the form
\begin{multline}
-\frac{dE}{dz}\Big|_{\rm col}
= n_e\Bigg[
2\pi A \ln\!\frac{R_{\max}}{R_{\min}}
-\frac{2\pi^2 e^2}{v}\!\int_{R_{\min}}^{R_{\max}}\!|\psi(R)|^2\,dR \\[4pt]
+\,2\pi C\Big(\cdots\Big)_{R_{\min}}^{R_{\max}}
\Bigg] .
\label{eq:full-dEdz}
\end{multline}

and setting $|\psi|^2\!\to0$ and $g_a\!\to0$ reduces Eq.\eqref{eq:full-dEdz} to a pure logarithm with a compactification-dressed prefactor $A$.

In anisotropic media, as neither the flux $T_{r0}$ nor the screening scale are azimuthally uniform, letting
\begin{equation}
2\pi \;\longrightarrow\; \int_0^{2\pi}\! d\phi,
\qquad
R_{\max}\to R_{\max}(\phi)\equiv \frac{\gamma v}{\bar\omega(\hat{\mathbf{n}}(\phi))},
\label{eq:anisotropic}
\end{equation}
modifies Eq.\eqref{eq:master-int} into
\begin{equation}
-\frac{dE}{dz}\Big|_{\rm col}
= n_e\!\int_0^{2\pi}\!\!\! d\phi
\int_{R_{\min}}^{R_{\max}(\phi)}\!
R\,\mathcal{I}(R,\phi)\,dR,
\end{equation}
where $\mathcal{I}$ represents the angularly resolved version of Eq.\eqref{eq:kernel-ABC} including the $\varphi^{ij}_a(\phi)$ contractions. In this case, Eq.\eqref{eq:yukawa-closed} applies inside the $\phi$–integral with the replacement $R_{\max}\to R_{\max}(\phi)$.

\subsection{Comparison with Linear Response Theory and Limiting Cases}
In linear response, fixing $R_{\max}=\gamma v/\bar\omega$ maps the first term of Eq.\eqref{eq:full-dEdz} to the familiar logarithmic behavior of the dielectric approach. The second and third terms, however, have no counterpart in a static scalar dielectric and capture short-range structure and mediator-induced anisotropy beyond Bethe. The approach used in this work thus provides a covariant, extensible route to collisional loss in media with microscopic anisotropy, nonlinearity, or extra-dimensional dressing.

When $M\!\to\!\infty$, the last term of Eq.\eqref{eq:full-dEdz} collapses to a small-$R$ correction absorbed into $\Delta_{\rm shell}$.  Nevertheless, for the case when $M\!\to\!0$, the last term of Eq.\eqref{eq:full-dEdz} merges into a logarithm, modifying the effective argument while leaving the structure of Eq.\eqref{eq:full-dEdz} intact.  Also, the Belinfante symmetrization guarantees $\partial_\mu T^{\mu\nu}=0$ on shell, indicating that the surface integral leading to Eq.\eqref{eq:master-int} is gauge invariant.

\subsection{Reduction to Bethe-Møller}
If one wishes to recover the Bethe-Møller formula \cite{Bethe1932,Moller1932,Sternheimer1952,Berger1984} from Eq.\eqref{ener}, the gauge coupling $g_4$ should satisfy $g_4^2 = \frac{3\pi^2 m_e v}{4\gamma^3 R_c^2 e^2 \Lambda_{\text{EM}}}$
for $R_c \sim \Lambda_{\text{EM}}^{-1}$ and $\Lambda_{\text{EM}} \gg m_e$ which, in the large $R$ limit, takes the explicit form
\begin{equation}
-\left.\frac{dE}{dz}\right|_{\text{col}}
\approx
\frac{2\pi e^4 n_e}{m_e v^2} 
\ln\left(\frac{\gamma^2 m_e v^2}{I}\right),
\label{BetheMoller}
\end{equation}
for $R_{\min}\!=\!1/(\gamma m_ev)$ and
$R_{\max}\!=\!\gamma v/I
$. This matches the leading-order  Bethe-Møller formula for electrons in natural units.

\section{Data Comparison Strategy}

To have a comparison between the generalized stopping power with tensor corrections and experimental data, one may use
\begin{equation}
-\frac{dE}{dz}\Big|_{\rm eff}
= K \,\ln\!\frac{R_{\max}}{R_{\min}} \;+\;\delta_{\rm Yukawa}(M,g_a),
\end{equation}
with $\Delta_{\rm shell}$ and $g_4$ absorbed into a single effective normalization constant $K$. This constant can be fitted at a reference energy, implying $\delta_{\rm Yukawa}$ would then control the predicted deviations across energy and direction. This fitting strategy mirrors the use of effective mean excitation energies in Bethe–type analyses, which have been extended to include tensor-mediated corrections.

\section{Conclusion}

To sum up, determining collisional stopping power from the inspection of the energy–momentum tensor obtained via an effective action leads to a framework that, in the appropriate limits, reduces to the familiar logarithmic Bethe–Møller structure,  while systematically incorporating additional short-range overlap corrections and tensor-mediated Yukawa contributions. Such an approach may provide an alternative route to modeling anisotropy, nonlinearity, and medium-specific excitations that lie beyond the reach of static dielectric approaches. 

From Eq.\eqref{son}, it becomes evident that if one were to examine the collisional stopping power via the steps followed in this study, three regimes would be required to be considered. At large impact parameters ($R \gg 1/m_e$), the first term in Eq.\eqref{ener} reproduces the Bethe-like logarithmic dependence, although it is modified by extra-dimensional physics through the effective coupling $g_4$ and Lorentz factor $\gamma$. On the other hand, in the quantum regime where $R \sim 1/m_e$, the fermionic term becomes significant, sensitive to the overlap between the wavefunction of the projected electron and the electronic states of the medium. For $R \lesssim M_a^{-1}$, the microstructure of the medium generates Yukawa-type terms producing resonant enhancements once $R \approx M_a^{-1}$ is satisfied. This multi-scale description directly correlates the stopping power with both fundamental parameters of the higher-dimensional theory and measurable material properties encoded in $\Phi_a^{\mu\nu}$, offering the potential to reveal certain experimental aspects of crystalline materials or engineered heterostructures with controlled $M_a$ scales and beyond. Furthermore, the construction is manifestly gauge invariant, exhibits the correct limiting behavior as $M\!\to\!\infty$ and $M\!\to\!0$, and can be matched to experimental stopping data by absorbing compactification effects and shell contributions into a single normalization constant. In this way, the framework not only reproduces the Bethe–Møller limit but also predicts direction and material-dependent deviations that could be tested in anisotropic crystals, layered heterostructures, or externally driven media. 

This study can then be treated as a higher-dimensional approach to collisional energy loss, complementing conventional dielectric-response methods and holding the potential to open a path toward incorporating tensor responses, finite-temperature dynamics, and extra-dimensional effects into a quantity of direct experimental relevance. Future work will focus on explicit parameter fits to stopping-power datasets and on extending the analysis to dynamical backgrounds and collective excitations.

\bibliography{reff.bib}
\bibliographystyle{apsrev4-2}

\end{document}